# Accurate early detection of Parkinson's disease from SPECT imaging through Convolutional Neural Networks

R. Prashanth

***Abstract* - Early and accurate detection of Parkinson's disease (PD) is a crucial diagnostic challenge carrying immense clinical significance, for effective treatment regimens and patient management. For instance, a group of subjects termed SWEDD who are clinically diagnosed as PD, but show normal Single Photon Emission Computed Tomography (SPECT) scans, change their diagnosis as non-PD after few years of follow up, and in the meantime, they are treated with PD medications which do more harm than good. In this work, machine learning models are developed using features from SPECT images to detect early PD and SWEDD subjects from normal. These models were observed to perform with high accuracy. It is inferred from the study that these diagnostic models carry potential to help PD clinicians in the diagnostic process.**

## I. INTRODUCTION

Parkinson's disease is a progressive neurodegenerative disorder affecting millions of people worldwide and it is characterized by the loss of dopaminergic neurons in the substantia nigra [1, 2]. The prevalence of PD is such that it affects around 1% of all population above age 60 and this prevalence increases with age [3]. The clinical diagnosis of PD is difficult as there are no definitive diagnostic tests and the diagnosis is based on the presence of cardinal symptoms, such as tremor at rest, rigidity, bradykinesia, and based on the response to PD medications [1]. But by the time, the patient manifests these symptoms, the patient most likely would have crossed the early stage of the disease [4]. Early detection of Parkinson's disease (PD) is an important clinical problem as earlier the diagnosis, earlier any appropriately targeted therapies could be initiated before any drastic deterioration [5]. It can also help develop treatments and identify patients eligible for therapeutic clinical trials [5].

SPECT imaging using $^{123}$I-Ioflupane (DaTSCAN or [123I]FP-CIT) have shown to increase the diagnostic accuracy of PD, mainly in the earlier stages of the disease, by showing the functional deterioration or dopaminergic deficit in the striatal region of the brain (which is one of the primary regions getting affected in PD) [6-9]. The accuracy of diagnosis of PD at an early phase is the poorest based on clinical indices as the early symptoms are mild/moderate, unlike in the advanced stages of the disease [4, 5]. Also these symptoms are common in other neurodegenerative disorders like essential tremor and multiple system atrophy, which often leads to misdiagnosis [10-12]. The effects of misdiagnosis are severe as it may lead to unnecessary medical examinations and therapies, and associated side-effects. Recent studies have shown that around 3.6% to 19.6% of clinically diagnosed PD do not show any dopmaminergic deficit, and these subjects are classified as being SWEDDs (Scans Without Evidence of Dopaminergic Deficit) [10-12]. Subsequent follow-up on these subjects have shown that they neither deteriorate nor respond to levodopa, and that their SPECT scans remain normal in the follow-up imaging. Thus, these subjects were considered highly unlikely of having PD and that the initial diagnosis of PD was incorrect [13-15]. These studies evidently point out that dopaminergic imaging is highly useful and that an abnormal imaging, at least in cases of diagnostic uncertainty, is strongly supportive of a diagnosis of neurodegenerative Parkinsonism (PS) such as PD.

In clinical practice, SPECT images are usually analysed by visual inspection and/or by region of interest (ROI) analysis [16]. Visual analysis relies on the judgment of the observer that heavily depends on his expertise, experience and knowledge [17]. ROI techniques involve outlining or positioning the ROI over the striatum (target region) and the occipital cortex (reference region), and a quantitative measure termed the background subtracted striatal uptake ratio is computed [6]. Despite odds, the latter method or the quantitative method is the most acceptable one, since, according to a trial study, it provides an excellent intra- and inter-observer agreement [18]. However, the ROI based approach relies on manual intervention for placing the ROIs.

There have been many studies that make use of machine learning techniques to develop predictive models from SPECT imaging features for the early detection of PD [11, 19-31]. Segovia *et al.* extracted voxels corresponding to the striatum and performed data decomposition using partial least squares followed by classification into controls and PS by means of a Support Vector Classifier (SVM) classifier [28]. Illan *et al.* also used voxels corresponding to the striatum to train a SVM classifier with linear kernel to classify controls and PS [29]. Rojas *et al.* used voxels corresponding to the striatum and then carried out feature reduction through principal component analysis (PCA) followed by classification using SVM [30]. Towey *et al.* performed feature extraction on all voxels through singular value decomposition followed by classification into PS or non-PS [31]. Huertas-Fernández *et al.* calculated the bilateral caudate and putamen uptake and asymmetry indices from SPECT images and developed predictive models using logistic regression, SVM and LDA to classify PD from vascular parkinsonism [32]. Kim *et al.* used image augmentation to increase the size of data and a classifier based on the Inception v3 model that can classify normal from abnormal SPECT scans [20].

There are also many studies using the SPECT data from the Parkinson's Progression Marker Initiative (PPMI), which is among the most popular, widely used and large database for PD. The same database is used in this paper as well to develop machine learning models for classifying PD [11, 19, 21-27, 33, 34]. Choi *et al.* trained a Convolutional Neural Network (CNN), which they called PD Net, using SPECT images to classify PD from normal and non-parkinsonism tremor [11]. They also used the model to classify SWEDD subjects. In their analysis, they had used the complete volume data, rather than considering a selected range of slices, due to which the CNN network became complex with many layers. Martinez-Murcia *et al.* also used a CNN to differentiate PD from others (healthy normal and SWEDD). They used a threshold based approach to select sub-volumes from the volume which they later input to the CNN). They observe that due to this sub-volume selection, the complexity of the CNN became small with just two convolutional layers [22]. Martínez-Murcia *et al.* used the features extracted from SPECT images through independent component analysis (ICA) to train a SVM classifier to distinguish PD from normal. They observe much better performance than their previous work using voxel-as-features approach [21]. Hirschauer *et al.* used data from different clinical examinations and SPECT imaging, and trained a Enhanced probabilistic neural network (EPNN) model to differentiate PD from SWEDD [19]. Oliveira *et al.* (2015) used voxels as features that were extracted based on volumes of interest defined (which required manual intervention) and an SVM classifier was used to classify PD from normal [23]. Oliveira *et al.* (2018) used the standard binding potential features along with other features related to the volume and length of the striatal region from SPECT images to train a SVM classifier that could classify PD from healthy normal [33]. Ortiz *et al*. extracted features from isosurfaces computed from the regions of interest and trained a CNN based model to classify PD from healthy normal [24]. Prashanth *et al.* (2017) computed shape- and surface fitting-based features and used machine learning methods to develop classification models to differentiate scans with deficit, as in PD, from scans without deficit, as in normal and SWEDD [27]. Prashanth *et al.* (2016) used data from multiple modalities including clinical examinations, laboratory examinations and dopaminergic imaging, and developed classification models to distinguish early PD from normal [26]. Prashanth *et al.* (2014) used the striatal binding ratios to develop classification and prognostic models for PD [25]. Zhang *et al.* used multimodal data which included SPECT imaging data to identify different PD subtypes through Long-Short Term Memory (LSTM) networks and Dynamic Time Warping (DTW) [34].

Few main limitations (combined) from these studies are as following: few needed manual intervention, few had lower diagnostic accuracy or used smaller dataset used for modelling and testing, few used the complete image volume rather than using only relevant slices (leading to more complex machine learning models) and few considered only two classes in their studies which are PD and normal.

In this work, machine learning techniques, mainly the CNN is leveraged, which inherently carries out feature extraction along with feature reduction, to develop predictive models that can classify PD from normal as well as capable of detecting SWEDD. Bayesian hyperparameter optimization is used to select an optimized and much more compact CNN architecture as compared to the networks used in literature. The SPECT data for the three groups, namely healthy normal, early PD and SWEDD, from the PPMI database is used. Only relevant slices from the SPECT volume are considered which helps significantly in preventing overfitting.

## II. Materials and Methods

### *A. Dataset details*

The data used in the study is from the Parkinson's Progression Markers Initiative (PPMI) database (http://www.ppmi-info.org/data). For up-to-date information, please visit http://www.ppmi-info.org. The PPMI is a landmark, large-scale, comprehensive, observational, international, multi-center study that recruits *de novo* (early-untreated) PD patients, and age and gender matched healthy normal subjects to identify PD progression biomarkers [4, 35].

In this work, data from 209 healthy normal, 443 early PD and, 80 SWEDD were considered. All the PD patients were in their early stages (Hoehn and Yahr (HY) [36] stage 1 or 2 with mean ± SD as 1.50 ± 0.50) and all the SWEDD subjects (these were the newly diagnosed PD patients based on clinical symptoms, but show normal dopaminergic imaging) also showed early stage (mean ± SD HY stage as 1.46 ± 0.53) PD symptoms. Table I shows the age, gender and HY stage distribution of subjects in the three groups. For the study, scans from the screening visit was considered, except for one normal subject for which the scan for the month 12 was considered, thus making the total number of samples in the normal class as 210. All the subjects in 3 groups are age and gender matched.

### *B. Image Preprocessing by PPMI*

All the SPECT scans taken at different PPMI sites undergo a standard pre-processing procedure before they are publically shared via the database [35]. This pre-processing is carried out so that all scans were in the same anatomical alignment (spatially normalized). The process includes reconstruction from raw projection data, attenuation correction, followed by applying a standard Gaussian 3D 6.0 mm filter, and then normalizing these images to standard Montreal Neurologic Institute (MNI) space. These pre-processed scans are then shared for public access and are the ones used for this analysis. The analysis pipeline is as shown in Fig 1.

TABLE I
Details of the subjects in terms of age, gender and the HY stage

| | Normal | | Early PD | | | SWEDD | | |
|---|---|---|---|---|---|---|---|---|
| | Count | Age (mean) | Count | Age (mean) | Count | Age (mean) | Count | Age (mean) |
| Female | 73 | 59.32 | 157 | 60.91 | 1.46±0.50 | 30 | 58.16 | 1.4±0.50 |
| Male | 136 | 61.65 | 286 | 62.13 | 1.53±0.50 | 50 | 61.80 | 1.5±0.54 |
| All | 209 | 60.79 | 443 | 61.7 | 1.51 ± 0.50 | 80 | 60.43 | 1.46 ± 0.53 |

*scan for one of the female subjects for month 12 also included in the study. HY stands for Hoehn and Yahr stage

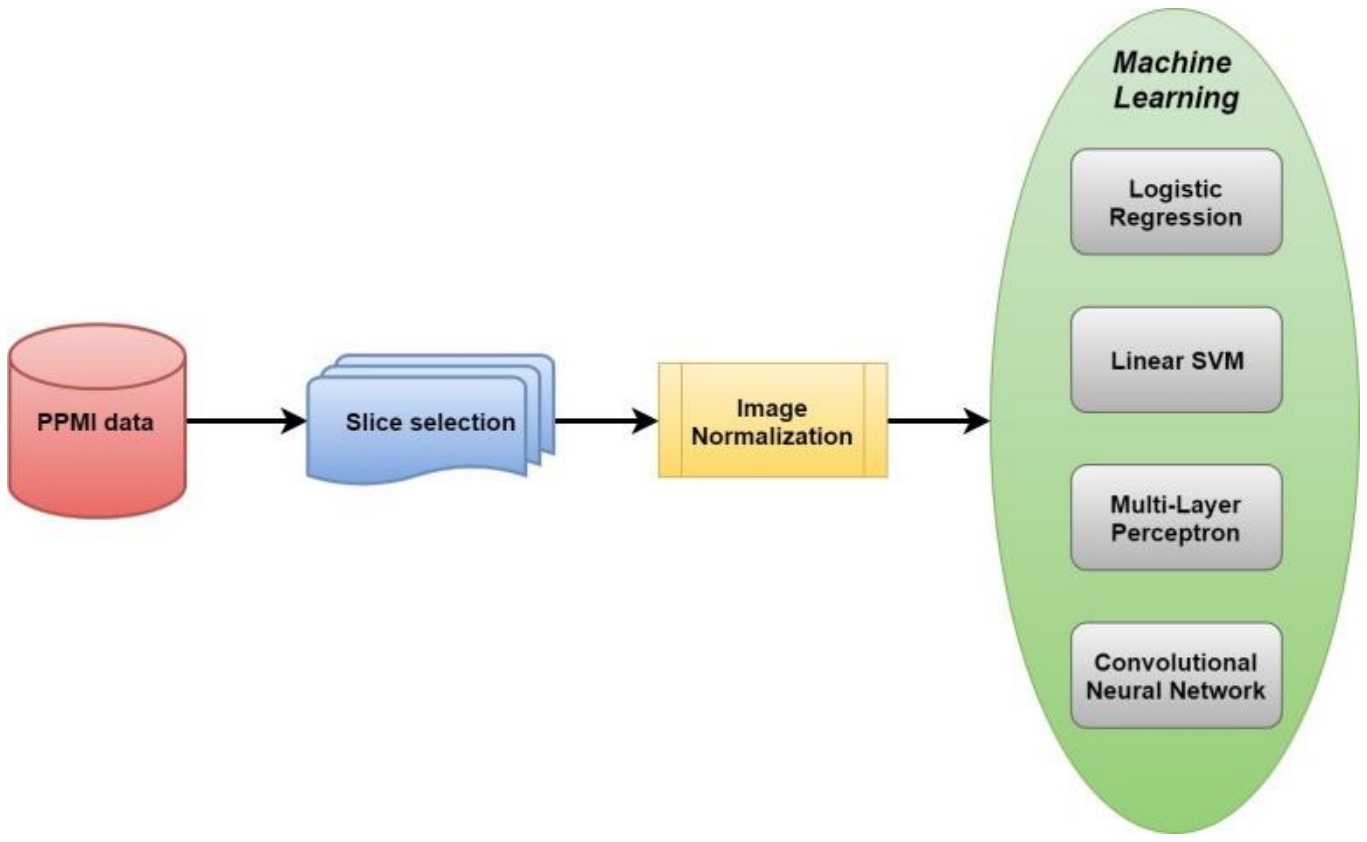

Fig. 1 Flowchart of the analysis

## C. *Slice selection*

Each SPECT scan consists of 91 transaxial slices (from bottom to top of head) each of size 109 x 91, which means each scan is of size 91 x109 x 91. In [27], the areas of striatal activity from SPECT images were computed and it was observed that the most relevant striatal activity was observed in slices from 35 to 48, with the highest activity in the slice number 41. In this work, two kinds of images are used for the analysis, one is the average of slices from 35 to 48, and the other is the 41$^{st}$ slice image only which is the slice with maximum striatal uptake. The idea of considering single slice and an average slice is because although the single slice image clearly shows the striatal activity, the average slice in a way aggregates the information from the whole of striatum. Figure 2 shows an illustration of both the average and single slice for the 3 groups which are Normal Control, Early PD and SWEDD. Normal scans are characterized by intense, uniform and symmetric high uptake or high intensity regions (corresponding to the caudate and striatum) on both hemispheres that appear as two 'comma' shaped regions, which are evident from Fig 2A and 2C. And in PD, deterioration of the dopaminergic neurons occur due to which the comma shaped region deteriorates, and becomes smaller and more circular in nature as observed in Fig 2B.

## D. *Image normalization*

The intensities in the original SPECT image ranged from 0 to $2^{15}$-1. The selected slice (the averaged one or the 41$^{st}$ slice) is normalized by dividing by $2^{15}$-1, so that the maximum intensity in the image could be 1 and the minimum could be 0.

| Averaged image | Single slice |
|---|---|
| A. Control | |

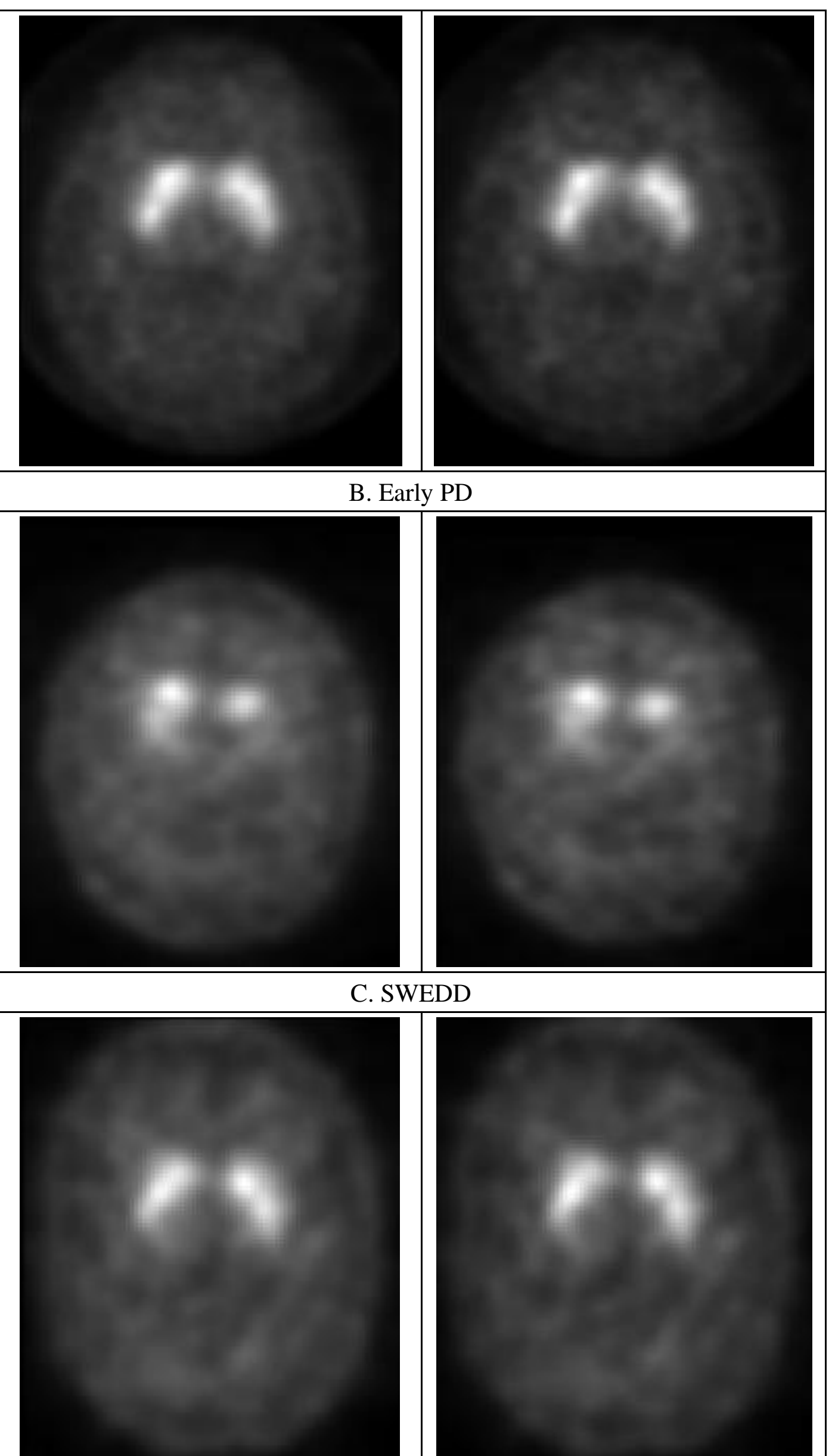

Fig. 2 An example case is shown for all the 3 groups. (A)Healthy normal control, (B) Early PD, (C) SWEDD.

## E. *Machine learning for early detection of PD*

Machine learning approaches were applied to develop predictive models that could differentiate early PD from other groups which are healthy normal and SWEDD. Ten fold cross validation was applied to evaluate the performance of the classifiers. The techniques used in the study include deep learning technique of Convolutional neural networks (CNN) [37], multilayer perceptron [37], support vector machine (SVM) [38] and logistic regression [39]. The normalized image is input to the methods to classify early PD from healthy normal. In the case of logistic regression and SVM, regularization is applied through the penalty parameter, and for MLP and CNN Bayesian optimization is used.

### 1) *Convolutional neural networks for predictive modelling*

In this work, a CNN is designed to classify early PD from healthy normal. CNNs are known to extract or compute higher-level representations of the image content, rather than carrying out preprocessing the data to derive features like textures and shapes. In other words, if a CNN is efficiently trained, it can avoid the step of feature extraction [37]. A CNN typically consists of a convolutional layer, a transformation, a pooling layer and a fully connected layer. In the convolutional layer, convolution operation occurs where tiles of the input feature map is extracted and filters are applied to them to compute new features. The parameters in this layer are the size of the tile and the number of filters. During the training, the CNN learns the optimal filter matrices that help in the extracting meaningful features (textures, edges, shapes) from the input feature map.

After the convolution operation, a transformation typically the Rectified Linear Unit (ReLU) is applied to the convolved feature. This will introduce nonlinearity into the model. After ReLU, a pooling step, typically max pooling is carried out. This will downsample the convolved feature, thereby reducing the dimensions of feature map while still preserving the most critical feature information. In max pooling, tiles are extracted and the maximum value is taken to generate new feature map. The parameters in the pooling layer are size of the max pooling filter and the stride which is the distance separating two consecutive extracted tiles, in pixels.

At last, there is fully connected layer which performs classification based on the features from the pooling layer. The parameters in a CNN model have to be fine tuned for optimal performance and to prevent over-fitting. For instance, the number of filters in the convolutional layer can be increased to get an increased number of features. However, more the filters, more resources will be used with increased training time. Additionally, each filter added may only provide insignificant incremental value than the previous one.

#### 2) *Fine tuning CNN – Hyperparameter optimization*

Using optimal parameters for the CNN is important for the best performance. Research has shown that Bayesian hyperparameter optimization of machine learning models especially neural networks is more efficient (with regard to overall performance on the test set and the time required to find the optimal parameters) than manual, random or grid search based methods [40]. In Bayesian optimization, unlike in random search, it keeps track of past evaluation scores which is used to form a probabilistic model mapping hyperparameters to a probability of a score on the objective function $p(y|x)$. Now this probabilistic model is much easier to optimize than the original objective function, thereby helping in finding the next best set of hyperparameters to evaluate. In our analysis, Tree-structured Parzen Estimator Approach (TPE) is used to estimate the probabilistic model [40]. The optimal architecture for CNN and MLP is estimated based on this optimization, and is given in the Results section.

Along with CNN, other machine learning methods including multilayer perceptron [37], support vector machine (SVM) [38] and logistic regression [39] were also used. Multilayer perceptron is a feed forward neural network with one or more hidden layers. The logistic model and SVM are the regularized ones with L1-normalization. This normalization is chosen because it can inherently carry out feature selection thereby reducing the number of features and improving numerical stability [39]. Regularizations are applied to MLP and CNN through Dropout [41].

## III. Results and Discussion

The parameters of classification algorithms (logistic regression and SVM) are estimated through cross validation. The optimal values for the regularization parameter were obtained as 1.0 and 0.5 for logistic regression and SVM, respectively. In case of MLP and CNN, parameters are estimated using Bayesian approximation. The optimized MLP model contained one hidden layer with 32 neurons with a dropout of 0.4 in the hidden layer. The optimized CNN model is shown in Table II below. The input goes to a convolution layer with 64 filters of size 5 x 5 with stride 1 and no padding, followed by a max pooling layer. Next there is another convolution layer with 32 filters of size 5 x 5 with stride 1 and no padding, followed by another max pooling layer. The output of this layer goes to a fully connected layer with 16 neurons with dropout as 0.2, followed by the final output layer of 2 neurons. This configuration is much more compact as compared to a related work [24] where they use 5 convolutional layers and 3 fully connected layers. And this compact configuration gives much better performance also. This result shows that optimization of CNN through Bayesian approximation is indeed helful.

TABLE II

THE ARCHITECTURE OF THE CNN MODEL

| Layer (type) | Output Shape |
|---|---|
| Input Layer | (109, 91, 1) |
| Conv2D (5 x5) | (105, 87, 64) |
| MaxPooling2D (2 x 2) | (52, 43, 64) |
| Conv2D (3 x 3) | (50, 41, 32) |
| MaxPooling2D (2 x 2) | (25, 20, 32) |
| Flatten | (16000) |
| Dense | (16) |
| Dropout (0.2) | (16) |
| Output Layer - Dense | (2) |

Table III below shows the 10-fold cross validation performance measures obtained using the average image and the single slice image. We observe excellent classification results for all the methods with CNN giving the best performance. It is also observed that the results between average image and single slice image are very close or similar. This might be because, as observed from Fig 2, the information content between averaged image and single slice image did not vary that much, with difference only in the intensity levels.

TABLE III

PERFORMANCE METRICS OBTAINED FOR DIFFERENT METHODS FOR (A) USING THE AVERAGE IMAGE, (B) USING THE SINGLE SLICE IMAGE

| Method | Confusion matrix | Accuracy | AUC | APR | Precision | Recall | Specificity |
|---|---|---|---|---|---|---|---|
| A. Average image | | | | | | | |
| Log Reg | $\begin{bmatrix} 431 & 12 \\ 12 & 198 \end{bmatrix}$ | 96.32% | 98.96% | 96.03% | 97.29% | 97.29% | 94.29% |
| LinearSVM | $\begin{bmatrix}429 & 14\\ 9 & 198\end{bmatrix}$ | 96.47% | 99.02% | 96.64% | 97.95% | 96.84% | 95.71% |
| MLP | $\begin{bmatrix}426 & 17\\ 5 & 202\end{bmatrix}$ | 96.63% | 98.73% | 95.72% | 98.84% | 96.16% | 97.62% |
| CNN | $\begin{bmatrix}433 & 10\\ 1 & 209\end{bmatrix}$ | 98.32% | 99.40% | 98.24% | 99.77% | 97.74% | 99.52% |
| B. Single slice image | | | | | | | |
| Log Reg | $\begin{bmatrix}431 & 12\\ 12 & 198\end{bmatrix}$ | 96.32% | 98.93% | 96.40% | 97.29% | 97.29% | 94.29% |
| LinearSVM | $\begin{bmatrix}429 & 14\\ 12 & 198\end{bmatrix}$ | 96.02% | 98.98% | 96.98% | 97.28% | 96.83% | 94.29% |
| MLP | $\begin{bmatrix}430 & 13\\ 8 & 202\end{bmatrix}$ | 96.79% | 99.48% | 98.77% | 98.17% | 97.07% | 96.19% |
| CNN | $\begin{bmatrix}439 & 4\\ 2 & 208\end{bmatrix}$ | 99.08% | 99.93% | 99.86% | 99.55% | 99.10% | 99.05% |

The confusion matrix is represented as $\begin{bmatrix}\text{True positive} & \text{False negative}\\ \text{False positive} & \text{True negative}\end{bmatrix}$. AUC stands for area under the region operating characteristic curve. LogReg, LinearSVM, MLP and CNN are the methods and stands for logistic regression, linear support vector machine, multilayer perceptron and convolutional neural networks, respectively

This work significantly improves the results obtained from [27] and other closely related works [11, 19-31, 33, 34]. In [27], classification model was developed for the detection of early PD from normal controls using features extracted from SPECT images, the best performances obtained was accuracy of 97.29% and AUC (area under the ROC curve) of 99.26%. This work notably improves these metrics with accuracy of 99.08% and AUC of 99.93% for single slice image. It is also to be noted that there is no feature extraction step in this analysis, unlike in the previous works. It is the extraordinary ability of CNN to extract a variety of features through convolutions and pooling that is leveraged here.

### A. *Error Analysis*

Fig. 3 illustrates few misdetections from the CNN model for the single slice image. Fig. 3A shows an image which belonged to the normal class but detected as early PD. It is to be noted that a normal scan is characterized by intense, uniform and symmetric high intensity regions on both hemispheres that appear as two 'comma' shaped regions (as observed in Fig. 2A). If we observe Fig. 3A, it is observed that tail or the bottom of the comma shaped region is less intense as compared to the upper region. This might be an interesting case of misdetection from the CNN model as the model actually is detecting the non-uniformity in the comma shaped region in the image. This can also happen that this might be a case of borderline, rather than a case of wrong labelling. Training the network with more images like these may help alleviating these errors. But in a way these errors could also help a clinician by indicating that these might be a case of borderline. Similarly, Fig. 3B is a case of misdetection where early PD case is detected as normal. Here as well, it is an interesting observation that the single slice appears normal (as the comma shaped region is clearly seen on both hemispheres). And the averaged image also appears similar (not shown here). But for the same subject, if we look at the slices individually from 35 to 48, there are changes in the intensity pattern in the comma shaped region and this might have been the reason for the subject to labelled as early PD. Giving weights to each slice and then averaging them can help improving the performance but again estimating a reliable weight for each image is a challenging problem.

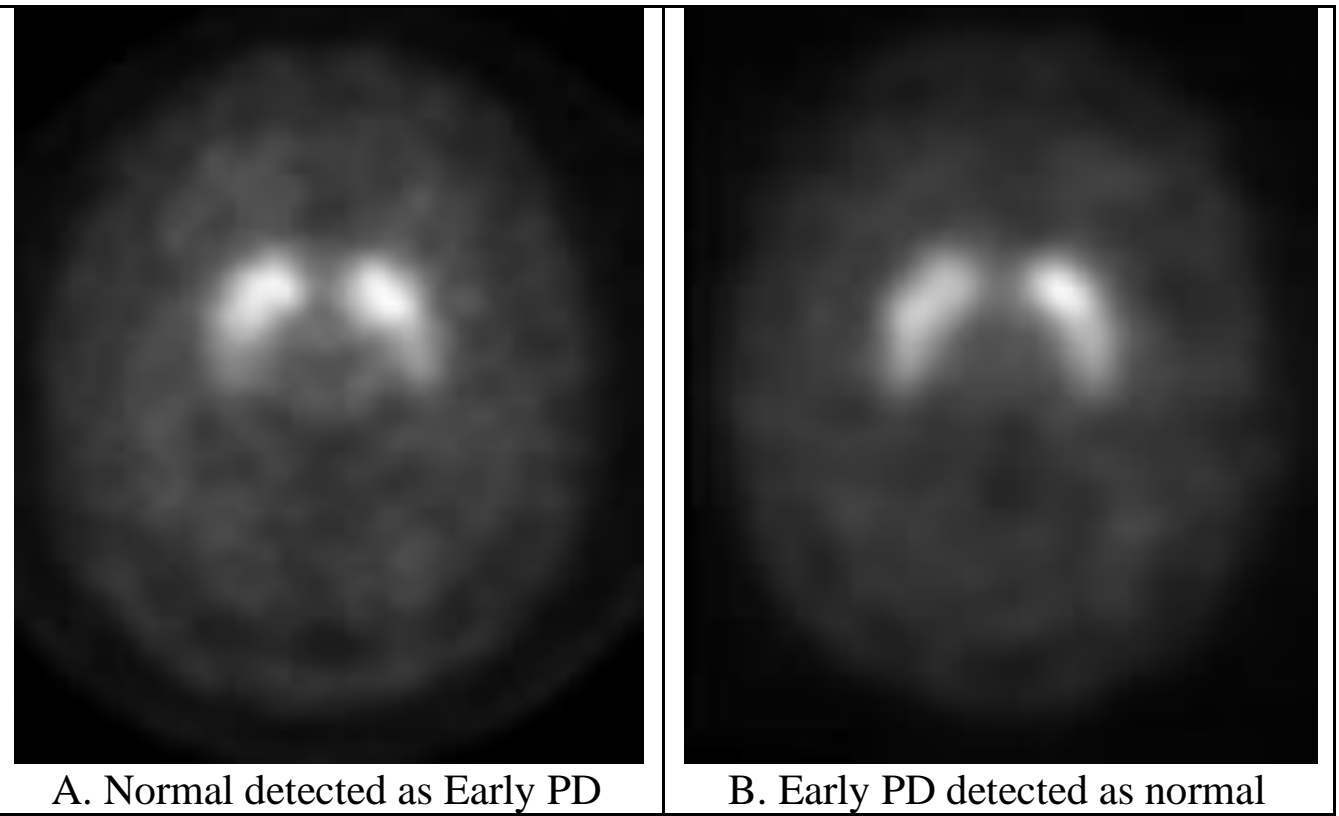

A. Normal detected as Early PD | B. Early PD detected as normal

Fig. 3 An illustration of misclassifications from the CNN model

### B. *Performance on SWEDD data*

The SWEDD data consist of 80 subjects and were input to the machine learned models. The performance of these methods is given is Table IV. CNN gave the best detection with accuracy of 95% (76 out of 80). Fig. 4 shows the cases of misdetection from the CNN model. It is interesting to observe that all these misdetected images show unexpected pattern which deviate from a normal behaviour as the comma shaped regions are uneven and dull. A recent study by Choi *et al* [11] which used the same PPMI data for analysis also observed that few SWEDD cases that showed unusual image pattern were detected as abnormal (or PD). And the diagnosis of the majority of these cases was later changed to clinical PD based on 2-year follow up. This shows the applicability of machine learning techniques here as these techniques, especially the CNN, could learn and infer using the training data.

TABLE IV
CLASSIFICATION RESULTS FOR THE SWEDD DATA FROM DIFFERENT METHODS

| | Averaged image | | Single slice image | |
|---|---|---|---|---|
| | True negative | False positive | True negative | False positive |
| CNN | 75 | 5 | 76 | 4 |
| Log Reg | 73 | 7 | 73 | 7 |
| LinearSVM | 73 | 7 | 73 | 7 |
| MLP | 74 | 6 | 73 | 7 |

### C. *Future work*

Recent research shows that deep learning techniques such as the CNN could benefit from the latest advances such as data augmentation which is a way to increase the training data using information from the available training data [42]. Traditional transformations which include a combination of various affine transformations and using Generative Adversarial Networks (GANs) [43] are effective ways to augment the data. Label smoothing is another advancement which has shown to improve the performance of deep learning models [44]. In label smoothing, the hard class labels are converted to soft labels. Both data augmentation and label smoothing are ways for regularizing the neural network models which can help in preventing overfitting and also help networks in converging faster.

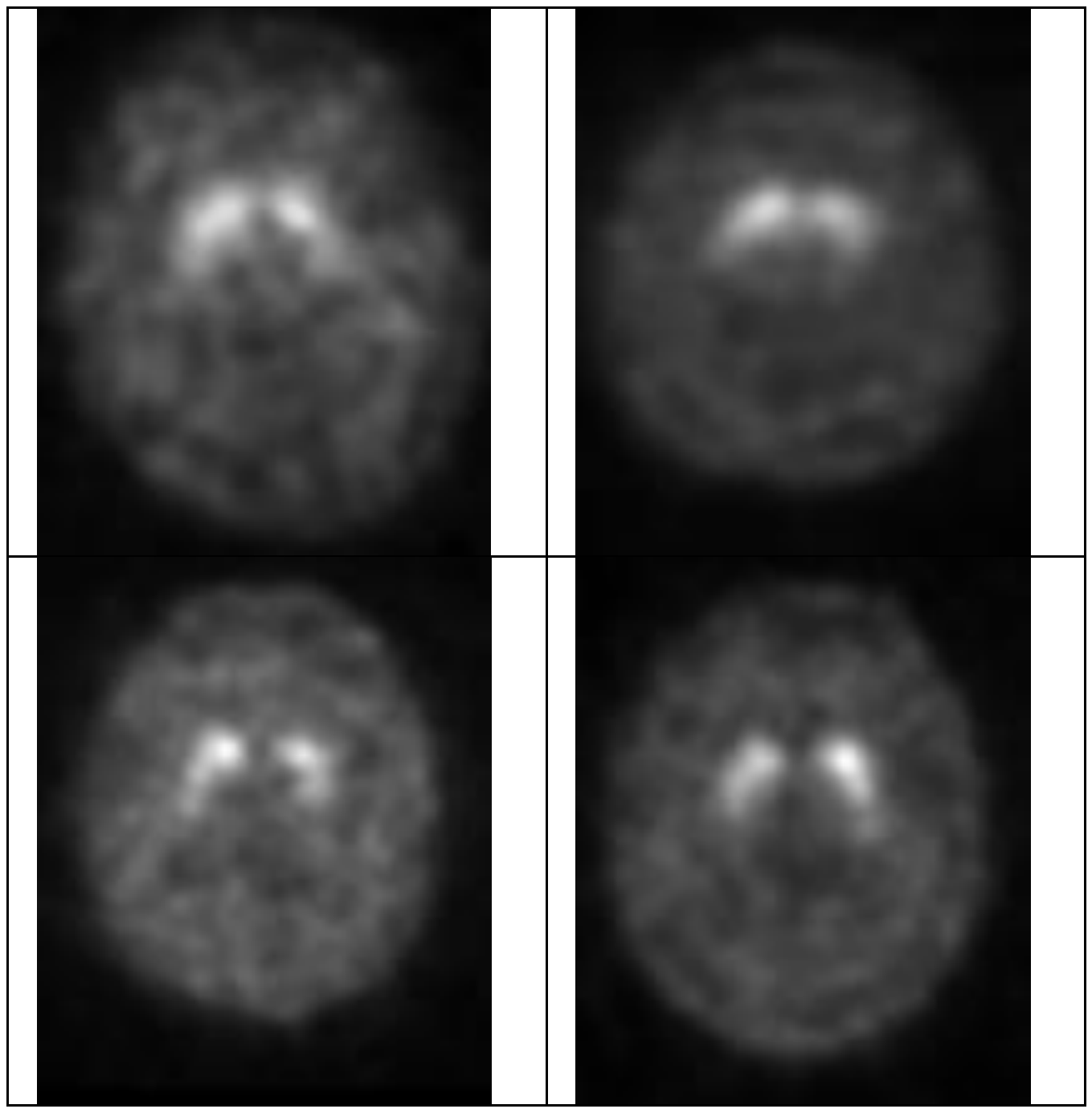

Fig. 4 Illustration of SWEDD images that were misclassified as early PDs by the CNN model

## IV. Conclusion

Accurate detection of PD from the non-degenerative ones (SWEDD) cases in their early stages is a challenging and important problem. As the class of parkinsonism disorders share many common symptoms, it is a source for misdiagnosis. Accurate identification of degenerative PS from other non-degenerative ones is crucial for effective patient treatment and management. In this work, machine learning models are developed that could classify subjects with early PD from healthy normal and also from SWEDD. These models gave good performance; especially the CNN model gave the most excellent performance among all methods achieving an AUC close to 100%. These predictive models carry enormous potential to be used in a clinical setting and can act as an aid to a clinician in the diagnostic process.

## Acknowledgment

PPMI, a public-private partnership, is funded by the Michael J. Fox Foundation for Parkinson’s Research and other funding partners include AbbVie, Allergan, Amathus Therapeutics, Avid Radiopharmaceuticals, Biogen Idec, BioLegend, Bristol-Myers Squibb, Celgene, Denali Therapeutics, GE Healthcare, Genentech, GlaxoSmithKline, Eli Lilly and Company, Lundbeck, Merck & Co., Meso Scale Discovery, Pfizer, Piramal, Prevail Therapeutics, Hoffmann-La Roche, Sanofi Genzyme, Servier, Takeda Pharmaceutical Company, Teva, Verily Life Sciences, Voyager Therapeutics, and UCB (Union ChimiqueBelge).